\documentclass[aip,apl,reprint,amsmath,amssymb]{revtex4-1}

\usepackage{graphicx}
\usepackage{txfonts}
\usepackage[dvips]{epsfig}
\begin{document}


\title{Bidirectional photon extraction from an epitaxially grown semiconductor quantum dot sandwiched by single mode optical fibers}

\author{H.\ Sasakura}
    \email[]{hirotaka@eng.hokudai.ac.jp}    
    	\affiliation{Research Institute for Electronic Science, Hokkaido University, Sapporo 001-0021, Japan}

\author{X.\ Liu}
    	\affiliation{Research Institute for Electronic Science, Hokkaido University, Sapporo 001-0021, Japan}

\author{S.\ Odashima}
\affiliation{Research Institute for Electronic Science, Hokkaido University, Sapporo 001-0021, Japan}

\author{H.\ Kumano}
       	\affiliation{Research Institute for Electronic Science, Hokkaido University, Sapporo 001-0021, Japan}
\author {S.\ Muto}
       	\affiliation{Department of Applied Physics, Hokkaido University, Sapporo 060-8628, Japan}

\author{I.\ Suemune}
\affiliation{Research Institute for Electronic Science, Hokkaido University, Sapporo 001-0021, Japan}

\date{\today}

\begin{abstract}
Fiber-based bidirectional photon extraction from nanoscale emitters and photon antibunching behavior between two outputs of two single mode optical fibers are experimentally demonstrated. Flakes of the epitaxial layer containing the InAs quantum dots (QDs) are fixed mechanically by both side with the edge faces of the single-mode-fiber (SMF) patch cables. The emitting photons from the single quantum dot are directly taken out of both side through the SMFs. Single-photon emission between two SMF outputs is confirmed by detecting non-classical antibunching in second-order photon correlation measurements with two superconducting single-photon detectors (SSPDs) and a time-amplitude converter (TAC). This simple opto-mechanical alignment-free single-photon emitter has advantage of robust stability more than 10 days and low-cost fabrication.

\end{abstract}
\maketitle

Semiconductor quantum dots (QDs) are attractive nanoscale structures for solid-state non-classical light sources and are expected to play key roles in quantum information network~\cite{Aws02}. Above all, single QDs fabricated by epitaxial growth can serve as stable and bright photon emitters. Recently, significant progress of far-field optical coupling to a single semiconductor QD has been achieved by the Purcell effects with elaborated microcavity structures~\cite{Moreau02,Claudon10,Reimer12,Takemoto07,Dousse10}. In addition, various efforts have been made to directly couple nanoscale photon emitter with conventional single mode optical fiber (SMF) from the perspective of consistency with existing optical fiber infrastructure. Several groups reported the efficient extraction of photons from a nanoscale emitter directly into optical fiber by using tapered fiber~\cite{Fujiwara11}, nanophotonic directional coupler~\cite{Davanco11}, and direct placing on the fiber facet by an atomic force microscope~\cite{Schroder11}. In these approaches, no optical loss occurs at coupling from the free-space into optical fiber. From a scientific and engineering point of view, the long term stability is one of the most important properties for fiber-based nanoscale photon emitter; however, the precise positioning, solid maintenance of geometry, and highly-accurate nanofabrication are required for the efficient coupling between the nanoscale emitter and the optical fiber. 

This study reports the fiber-based bidirectional solid-state single-photon emitter based on epitaxially grown InAs QDs using SMFs. The photons emitted from the single InAs QD as single-photon emitter are taken out of both side through the SMFs. Flakes of the epitaxial layer containing the InAs QDs are made by simply scratching surface with an ordinary diamond cutter and are fixed mechanically by both side with the edge faces of the SMF patch cables. Clear antibunching between two outputs of two single mode fiber patch cables is observed in second-order photon correlation measurements with superconducting single photon detectors and time amplitude converter. Although it is not a mainstream technique for the realization of high extraction efficiency from nanoscale photon emitter, placing the solid-state single-photon source and the the SMFs in direct contact is useful from a practical perspective [Figure 1(a)] because it offers robust stability and does not require opto-mechanical alignment. Moreover, this simple structure facilitates low-cost for fabrication of photon emitters and sufficient stability of output photon number from a scientific viewpoint.

\begin{figure}[h]
\includegraphics[width=20.5pc]{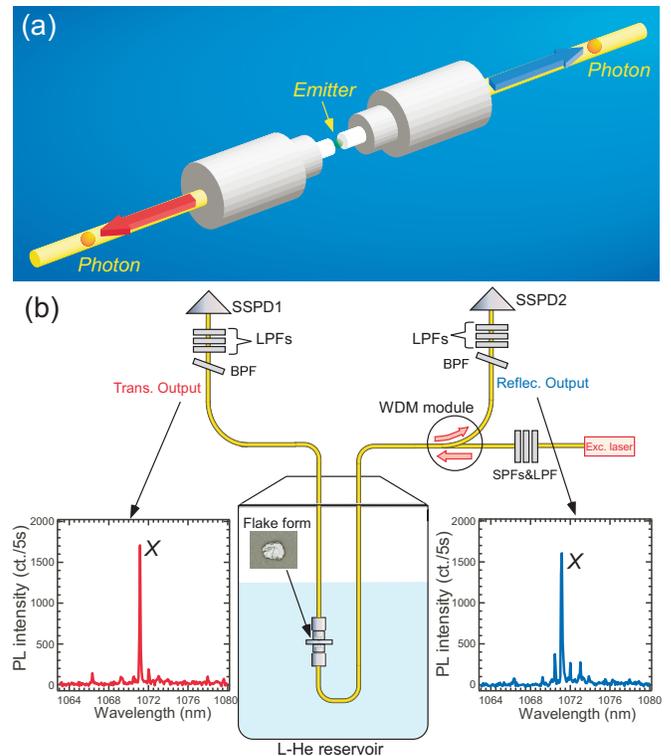}\hspace{2pc}%
\caption{\label{fig0} (Color online) Schematic of SMF embedded solid-state single-photon emitter (a) and experimental setup of second-order photon correlation (b). (Inset): Measured PL spectra in transmission (Trans. Output) and reflection (Reflec. Output) configurations. SPF is short-pass filter, LPF is long-pass filter, BPF is 0.5-nm-wide band-pass filter, and SSPD is superconducting single-photon detector.}
\end{figure}

The InAs QDs were grown to a density of 7$\times$$10^{9}$ cm$^{-2}$ on a GaAs buffer layer on a semi-insulating GaAs (001) substrate by molecular-beam epitaxy (RIBER, MBE32P). The low-temperature indium flush method was user to restrict the QD height to 2.8 nm. The detailed growth conditions and overall optical characteristics are presented in Ref.~\onlinecite{Sasa}.
 To extract emitted photons in both directions, we used flake forms of the epitaxial layer. The flakes having diameter and thickness of 1-20 $\mu$m were obtained by simply scratching surface with an ordinary diamond cutter. These flakes were directly attached to the edge face of a FC/PC SMF patch cable (Corning, SMF-28) with a $\phi$900 $\mu$m jacket by electrostatic forces. To mechanically fix the flakes and effectively extract the generated photons into two SMFs, the edge face of another SMF patch cable was directly connected to the opposite side of the flakes using a FC/PC-FC/PC joint (Thorlabs, ADAFC2). The FC/PC-FC/PC joint part with its embedded flakes was set in a liquid $^4$He reservoir at 4.2 K [Fig.~\ref{fig0}(b)].

We used a fiber-pigtailed laser diode (Thorlabs, LPS-830-FC) emitting at 830 nm as an excitation source with the laser beam arriving at the flakes from right side of the SMF. To obtain a clean laser spectrum, the long-pass and two short-pass filters were inserted into the laser beam (Thorlabs, FEL0800 and FES0850, respectively). The QD emissions were extracted using two SMFs, corresponding to the reflection (right side) and transmission (left side) configurations in the free-space optical measurement system. To spatially separate the emissions that traveled in the SMF in the direction opposite to that of excitation laser, a wavelength division multiplexing module (Optoquest, custom-made product) was used. The emission was dispersed by a double-grating spectrometer (Acton, Spectrapro 2500i, $f=1.0$ m) and detected with a liquid-nitrogen-cooled InGaAs photodiode array (Roper, OMA-V1024). The typical exposure time was 5 s to obtain a spectrum with a high signal-to-noise ratio. 

The sharp peaks originating from single InAs QDs were observed at the reflection and transmission configuration under the excitation power density of 5.6 W/cm$^2$ [Fig.~\ref{fig0}(b)]. At low excitation power, the emission $X$ is centered at 1071.12 nm with a 76 $\mu$eV full width at half maximum (hereinafter called the $X$ line), which corresponds to the ground-state emission in the ensemble photoluminescence (PL) spectra~\cite{Sasa}. The observed intensities of the $X$ line in each configuration were almost identical, signifying that the present system that uses flakes sandwiched between SMF patch cables naturally forms a Hanbury Brown and Twiss (HBT) interferometer. Note that the intensity difference between the side peaks at 1070.5 and 1073.0 nm in the observed spectra at right and left side is attributed to a slight difference of position of the single QD in the flake.

\begin{figure}[h]
\includegraphics[width=20.5pc]{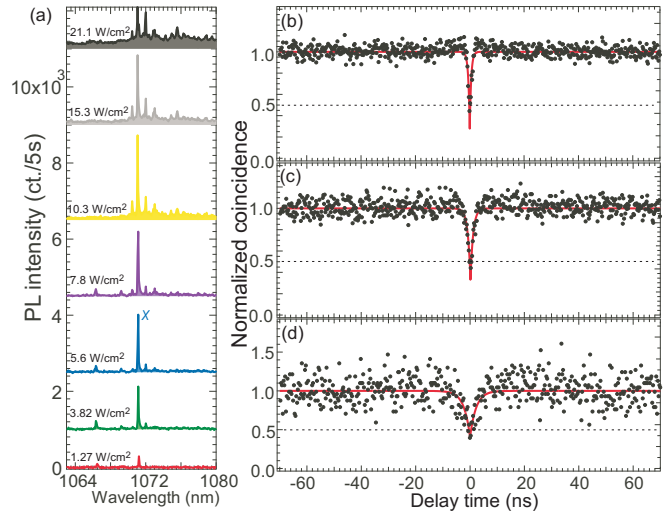}\hspace{2pc}%
\caption{\label{fig2} (Color online) (a) PL excitation power dependence. Photon count is displayed as a function of optical delay. Second-order photon correlation is recorded by a TAC employing two SSPDs. Normalized histograms of the autocorrelation measurement of the $X$ line (solid circles) at excitation power of 5.6 W/cm$^2$ (b), 3.82 W/cm$^{2}$ (c), and 1.27 W/cm$^2$ (d). Excitation wavelength is 830 nm and the time bins are 244 ps. The integration time and count rates are 10.5 h and $\sim$16 kHz (b), 7.5 h and $\sim$8 kHz (c), and 14.5 h and $\sim$3 kHz (d), respectively. The solid (red) lines are fits, indicating a dip in the correlation function of $g^{(2)}(0)=0.28$ and $t$=0.54 ns (b), 0.33 and 1.0 ns (c), and 0.41 and 2.24 ns (d). }
\end{figure}

Figure~\ref{fig2}(a) shows 
the excitation power dependence of the single InAs QD PL spectrum in the transmission configuration. At low excitation power the $X$ line is dominates, implying that the QD emitting the $X$ line couples optically well with the SMF core. With increasing excitation power, additional PL lines appear, which are exciton complexes of the QD and/or emissions originating from other QDs. At the highest excitation power density (21.1 W/cm$^{2}$), background photons appear, suggesting that a large number of QDs couple weakly with the SMFs. Although the flake with diameter $<$10 $\mu$m contains $<$5000 QDs, the present system corresponds to the standard micro-PL measurements in the far field using a collective lens with low numerical aperture. 

For the auto-correlation measurements, the output of each SMF is filtered with a 0.5-nm-wide band-pass filter (Optoquest: custom-made product) to select the $X$ line and three long-pass filters (Asahi Spectra, LIX900 and Thorlabs, EFL1000) to suppress background photons, i.e., reflection and transmission of the excitation laser at the interface of both SMF patch cables and other unwanted emissions originating from other QDs. The optical transmission loss of each part in selecting the $X$ line was $\sim$5 dB. The $X$ line was then sent to a superconducting single-photon detector (SSPD) (Single Quantum, custom-made product). The detection efficiency and dark count of each SSPD were $\sim$15\% and $\sim$50 cps, respectively~\cite{Dorenbos11}. The auto-correlation measurements of the $X$ line were performed between the right and left ports with an optical delay of $\sim$100 ns to compensate for the internal electronic delay in the SSPD driver and in the time-amplitude converter (TAC) board (Becker \& Hickl, SPC-130). The single count rate of the $X$ line at each SSPD was $\sim$8 kHz with an excitation power of 5.6 W/cm$^2$. A histogram of the normalized coincidence counts with time bins of 244 ps and an integration time of 10.5 h is shown in Fig.~\ref{fig2}(b). The data exhibits the well-known antibunching dip at zero time delay and is fit with the second-order correlation function $g^{(2)}(\tau)=1-\alpha \exp(-|\tau|/t)$, where $\alpha =0.72\pm 0.05$ accounts for the background light and dark counts of the SSPD and $t[=t_{l}t_{p}/(t_{l}+t_{p})]=0.54\pm 0.10$ ns, where $t_{l}$ is the lifetime of the $X$ line and $t_{p}$ is the inverse pumping rate. At zero delay, the function gives $g^{(2)}(0)= 0.28\pm 0.05$, suggesting that the single InAs QD emits non-classical light. Figures ~\ref{fig2}(c) and 2(d) show normalized coincidence under an excitation power of 3.82 W/cm$^{2}$ and 1.72 W/cm$^{2}$, respectively. Upon decreasing the excitation power, which corresponds to increasing $t_{p}$, the full width at half maximum of the antibunching dip gradually spreads up to $\sim$2.24 ns, suggesting that $t$ is not limited by $t_{l}$ in our measurement conditions. Although a narrow antibunching appears in Figures~\ref{fig2}(b)-2(d), a broad bunching extending over $\pm$10 ns near the narrow antibunching dip appears for an excitation power of 5.6 W/cm$^2$. This bunching photon statistics are suppressed under low-power excitation, which we attribute to an enhanced probability of the exciton state ($X$) re-pumping, after emitting a photon~\cite{Regelman01} or the spectral diffusion effects due to environmental fluctuations~\cite{Sallen10,Abbarchi12}.

In conclusion, we demonstrated single-photon emission from the flake containing epitaxially grown InAs quantum dots. The emitting photons originating from the single InAs quantum dot could be extracted bidirectionally through single mode optical fibers attached to the flake forms directly. The photon antibunching behavior was confirmed by second-order photon correlation measurements with the superconducting single photon detectors and time amplitude converter. These experimental results suggest that simple single-photon emitter based on the single quantum dot sandwiched by two single mode fiber patch cables has significant potential for robust stability of the extraction efficiency, opto-mechanical alignment-free, and fabricating low-cost photon emitters. Moreover bidirectional photon extraction approach has a potential to send a photon pair generated by nanoscale emitter to the separate directions without optical loss in coupling from free-space into optical fiber.

This work was supported in part by the Sumitomo foundation and Nikki-Saneyoshi foundation. X.L., H.K., and I.S. acknowledge founding from SCOPE.

\end{document}